\documentclass[amsmath, amssymb, breqn, aps, prd, superscriptaddress, twocolumn, notitlepage, longbibliography]{revtex4-2}

\usepackage[english]{babel}
\usepackage[utf8]{inputenc}
\usepackage{amsthm}
\usepackage{mathtools}
\usepackage{physics}
\usepackage{xcolor}
\usepackage{graphicx}
\usepackage{adjustbox}
\usepackage{placeins}
\usepackage[T1]{fontenc}
\usepackage{lipsum}
\usepackage{csquotes}
\usepackage{float}
\usepackage{units}
\usepackage{tabularx}
\usepackage{booktabs}
\usepackage{bm}
\usepackage{natbib}
\usepackage[caption=false,labelformat=empty,captionskip=0pt]{subfig}
\usepackage[colorinlistoftodos, color=green!40, prependcaption]{todonotes}
\usepackage[hidelinks,pdftex,pdftitle={Article},pdfauthor={Author}]{hyperref} 

\begin{document}
\title{HeLIOS: The Superfluid Helium Ultralight Dark Matter Detector}
\author{M. Hirschel}
    \affiliation{Department of Physics, University of Alberta, Edmonton, AB T6G 2E9, Canada}
\author{V. Vadakkumbatt}
    \affiliation{Department of Physics, University of Alberta, Edmonton, AB T6G 2E9, Canada}
\author{N.P. Baker}
    \affiliation{Department of Physics, University of Alberta, Edmonton, AB T6G 2E9, Canada}
\author{F.M. Schweizer}
    \affiliation{Department of Physics, University of Alberta, Edmonton, AB T6G 2E9, Canada}
\author{J.C. Sankey}
    \affiliation{Department of Physics, McGill University, Montr\'eal, Qu\'ebec H3A 2T8, Canada}
\author{S. Singh}
    \affiliation{Department of Electrical and Computer Engineering, University of Delaware, Newark, DE 19716, USA}
\author{J.P. Davis}
    \affiliation{Department of Physics, University of Alberta, Edmonton, AB T6G 2E9, Canada}
    
\begin{abstract}
    The absence of a breakthrough in directly observing dark matter (DM) through prominent large-scale detectors motivates the development of novel tabletop experiments probing more exotic regions of the parameter space. If DM contains ultralight bosonic particles, they would behave as a classical wave and could manifest through an oscillating force on baryonic matter that is coherent over $\sim 10^6$ periods. Our Helium ultraLIght dark matter Optomechanical Sensor (HeLIOS) uses the high-$Q$ acoustic modes of superfluid helium-4 to resonantly amplify this signal. A superconducting re-entrant microwave cavity enables sensitive optomechanical readout ultimately limited by thermal motion at millikelvin temperatures. Pressurizing the helium allows for the unique possibility of tuning the mechanical frequency to effectively broaden the DM detection bandwidth. We demonstrate the working principle of our prototype HeLIOS detector and show that future generations of HeLIOS could explore unconstrained parameter space for both scalar and vector ultralight DM after just an hour of integration time.
\end{abstract}

\maketitle

\section{Introduction}

While numerous astrophysical observations support the existence of dark matter (DM) \cite{Tyson1998,Markevitch2004,Planck2020}, its first direct detection is still awaited and is one of the greatest ambitions in modern science. Tremendous efforts toward detecting popular DM candidates like weakly interacting massive particles (WIMPs) have not been successful yet, motivating a stronger focus on more exotic regions of the DM parameter space, with a possible mass range spanning approximately 90 orders of magnitude \cite{Hu2000,Carr2016}. Extending the search into these territories is accompanied by the need for novel detection paradigms, with a growing focus on small-scale quantum systems as precise detectors for lowest-mass DM candidates \cite{Carney2021,Hochberg2022}.

Ultralight dark matter (UDM) denotes particle candidates with masses $m_\text{dm}<10~\text{eV}/c^2$ \cite{JacksonKimball2023,Antypas2022}. In this range, the local DM density $\varrho_\text{dm} = \unit[0.4]{GeV/cm^3}$ implies a bosonic particle occupation exceeding one in each de Broglie volume $\lambda_\text{dB}^3$, with $\lambda_\text{dB} = h/m_\text{dm} v_\text{vir}$ and particle velocity $v_\text{vir} \approx 10^{-3} c$ in the virialized DM halo \cite{Derevianko2018}. As a result, UDM particles would behave wave-like, i.e.,~could either be described through a classical (pseudo)scalar field $\Phi(\bm{x},t) \approx \Phi_0 \cos(\omega_\text{dm} t - \bm{k}_\text{dm} \cdot \bm{x})$ (for spin $S = 0$), or vector field $A_\mu(\bm{x},t) \approx A_{\mu,0} \cos(\omega_\text{dm} t - \bm{k}_\text{dm} \cdot \bm{x})$ (for $S = 1$), oscillating at the Compton frequency $\omega_\text{dm} = m_\text{dm} c^2/\hbar$ with wavevector $|\bm{k}_\text{dm}| = 2 \pi/\lambda_\text{dB}$ \cite{Arvanitaki2015,Stadnik2015a,Stadnik2015b,Derevianko2018}. Significant efforts toward UDM detection have been devoted to axions \cite{Sikivie2021} --- pseudoscalar particles in the $\text{\textmu}$eV-range --- for example through haloscopes like HAYSTAC \cite{HAYSTAC2017,HAYSTAC2021} or ADMX \cite{ADMX2018,ADMX2021b}.

A broad region of UDM parameter space below the $\text{\textmu}$eV-range has been excluded indirectly through astrophysical probes \cite{Hu2000,Marsh2016} and tests of equivalence principle violations \cite{VectorEotWash,ScalarEotWash}. Beyond these limits, mechanical sensors are predestined to directly search for UDM fields in the kilohertz range \cite{Arvanitaki2016,Carney2021}. The gravitational wave interferometers GEO600 \cite{ScalarGEO600} as well as LIGO and Virgo \cite{VectorLIGOVIRGO} provided broadband upper limits for scalar and vector UDM coupling by using the beamsplitter and mirrors as susceptible elements. On the other hand, mechanical resonators with low dissipation can provide amplification of the signal to overcome technical noise and achieve thermal-noise limited readout close to their mode frequencies, at the cost of a small bandwidth. Remarkable sensitivities have been reached in cryogenic systems \cite{Saulson1990}, such as the resonant-mass detector AURIGA, which set the strongest constraints on kilohertz scalar UDM to date \cite{ScalarAURIGA}. Several tabletop resonators have been proposed to effectively probe scalar \cite{Arvanitaki2016,Manley2020} and vector \cite{Manley2021} UDM parameter space. Cavity optomechanical systems provide an ideal platform to sensitively probe the motion of such a mechanical sensor, using a microwave or optical readout mode \cite{Aspelmeyer2014}. Optomechanical transducers with quantum-limited sensitivity have been envisioned in various proposals for sub-GeV \cite{ODIN2023,Carney2021,Afek2022} and UDM detection \cite{Carney2021,Manley2020,Manley2021,Brady2022,Murgui2023}.

\begin{figure*}[t]
\centering
    \subfloat[][]{\includegraphics[width=0.49\linewidth]{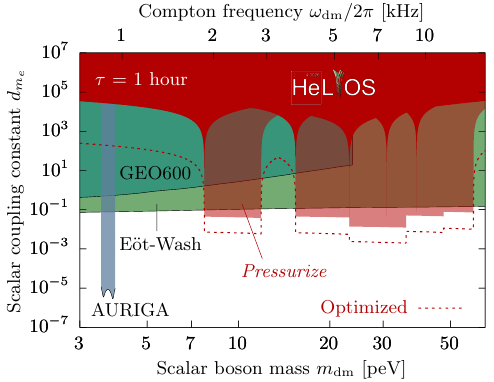}}\hfill
    \subfloat[][]{\includegraphics[width=0.49\linewidth]{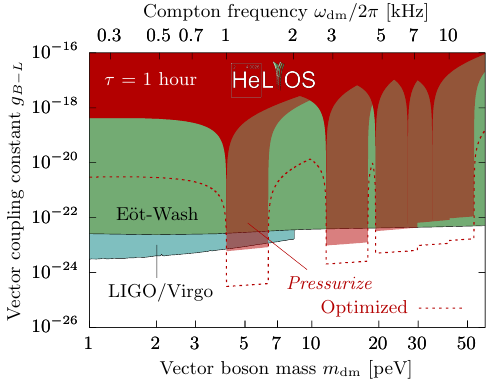}}
    \caption{Region of scalar UDM (left; coupling to the electron mass through $d_{m_e}$) and vector UDM (right; coupling to the baryon-minus-lepton number through $g_{B-L}$) parameter space accessible to the lowest ten modes of the current HeLIOS prototype after $\tau = 1 \, \text{hr}$ integration time (red), including frequency tuning through pressurization (transparent red). The dashed lines show limits achievable with realistic optimizations of the same design (using a temperature of $T = \unit[10]{mK}$, microwave power $P_\text{c} = \unit[10]{\text{\textmu}W}$, microwave mode decay rate $\kappa/2 \pi = \unit[1]{MHz}$ and frequency shift $\partial \omega_\text{c}/\partial P = - 2 \pi \times \unit[2]{GHz/bar}$ \cite{Vadakkumbatt2021}, mechanical $Q = 10^7$, as well as a doubled cylinder radius). Currently existing bounds by AURIGA \cite{ScalarAURIGA}, GEO600 \cite{ScalarGEO600}, LIGO/Virgo \cite{VectorLIGOVIRGO}, and the Eöt-Wash experiment \cite{ScalarEotWash,VectorEotWash} are shown in the background.}
    \label{fig:ParameterSpace}
\end{figure*}

Superfluid helium has been considered as a target material for particle-like sub-GeV DM in various proposals \cite{Lanou1987,Guo2013,Schutz2016,You2023,ODIN2023,Lyon2023} and has been used in experiments such as HERON \cite{HERON2017}, HeRALD \cite{HeRALD2019,HeRALD2023}, or DELight \cite{DELight2022}. For UDM searches, superfluid $^3$He \cite{Gao2022} and $^4$He \cite{Manley2020} have been proposed as promising detection media. Here, we introduce HeLIOS (Helium ultraLIght dark matter Optomechanical Sensor) --- the first UDM detector based on superfluid helium, capable of simultaneously searching for both scalar and vector bosons. Superfluid $^4$He is an ideal resonant mass for two reasons. First, it features ultra-low dissipation at millikelvin temperatures, with demonstrated mechanical $Q$ of more than $10^8$ \cite{deLorenzo2017}. Second, helium is the only element that remains liquid at low temperatures. Consequently, it can be pressurized to continuously change the speed of sound and tune the acoustic mode frequencies up to \unit[54]{\%}, thus overcoming the sensitivity-bandwidth trade-off and effectively enabling broadband detection \cite{Abraham1970,Vadakkumbatt2021}. Such tunability has already been recognized as a key detection feature, for example in the axion haloscopes \cite{ADMX2021a,HAYSTAC2020} that employ frequency-tunable high-$Q$ microwave cavities.

\section{Detecting ultralight dark matter with mechanical sensors}

In the standard halo model, DM has a Maxwellian velocity distribution \cite{Krauss1985}, leading to a Doppler shift and broadening of the UDM frequency. This limits the coherence time to $\tau_\text{dm} = 10^6/\omega_\text{dm}$ and coherence length to $\lambda_\text{dB}$, which exceeds \unit[1000]{km} for frequencies less than \unit[300]{kHz} \cite{Derevianko2018}. Thus, the UDM field can be considered coherent for a million periods and spatially uniform for any lab-scale experiment probing the kilohertz range.

Scalar UDM could linearly couple to standard model (SM) fields, effectively leading to a modulation of the fine-structure constant or fermion masses \cite{Derevianko2018}. This would result in a homogeneous strain $h(t) = - d h_0 \cos(\omega_\text{dm} t)$ imposed on any condensed body, with dimensionless coupling strength $d$ and amplitude $h_0$ \cite{Arvanitaki2016,Manley2020}. The oscillating strain acts as a driving force $F_\text{dm}(t) = q_n \Ddot{h}(t)$ when coupled to a narrow-band breathing mode $n$ of a mechanical resonator, with normalized mode shape $\tilde{\bm{u}}_n(\bm{x})$, effective mass $\mu_n = \int \rho \, |\tilde{\bm{u}}_n(\bm{x})|^2 \, \text{d}^3x$, geometric mode overlap factor $q_n = \int \rho \, \tilde{\bm{u}}_n(\bm{x}) \cdot \bm{x} \, {d}^3 x$, and mass density $\rho$ \cite{Hirakawa1973,Manley2020}.

On the other hand, vector UDM (or "dark photons") could couple to SM fields through dark charges of an object, such as the baryon number $B$ or baryon-minus-lepton number $B-L$ \cite{Carney2021}. In the center-of-mass frame, each object $j$ would experience an acceleration $\bm{a}_{j}(t) = f_{j} g \bm{a}_0 \cos(\omega_\text{dm} t)$, with material-dependent suppression factor $f_{j}$, dimensionless coupling strength $g$ and amplitude $\bm{a}_0$ \cite{Manley2021}. When acting on the mechanical modes of a resonator composed of two materials, the differential acceleration results in a driving force $\bm{F}_\text{dm}(t) = \mu_n [\beta_{n,1} \bm{a}_1(t) + \beta_{n,2} \bm{a}_2(t)]$, with geometric mode overlap factors $\beta_{n,j} = \int_j \hat{\bm{a}} \cdot \tilde{\bm{u}}_n(\bm{x}) \, \text{d}^3x / \int_{1 \cap 2} |\tilde{\bm{u}}_n(\bm{x})|^2 \, \text{d}^3x$ and acceleration polarization $\hat{\bm{a}} = \bm{a}_0/|\bm{a}_0|$ \cite{Manley2021}.

Ultimately, detection requires a signal-to-noise ratio larger than unity. The signal force power spectral density (PSD) $S_{FF}^\text{dm}$ for the respective UDM coupling (Eqs.~\eqref{eq:ScalarUDMPSD} and \eqref{eq:VectorUDMPSD}) as well as the considered noise contributions, adding up to the noise force PSD $S_{FF}^\text{noise}$ (Eq.~\eqref{eq:DetectorNoiseBudet_Appendix}), are discussed in the Appendix. Equating both gives an estimate for the force sensitivity, i.e.,
\begin{equation}
    \label{eq:DetectorNoiseBudet}
    S_{FF}^\text{dm}(\omega_\text{dm}) = S_{FF}^\text{noise}(\omega_\text{dm}) \sqrt{\frac{\tau_\text{dm}}{\tau}}
\end{equation}
for each normal mode (assuming no mode overlap). The factor $\sqrt{\tau_\text{dm}/\tau}$ accounts for the noise reduction when estimating the incoherent signal PSD over long integration times $\tau > \tau_\text{dm}$ through averaging $\tau/\tau_\text{dm}$ independent periodograms \cite{Budker2014,Manley2020,Manley2021}.

We consider scalar UDM coupling to the electron mass $m_e$ ($d = d_{m_e}$) and vector UDM coupling to the baryon-minus-lepton number $B-L$ ($g = g_{B-L}$). However, the same approach can also constrain coupling to the fine-structure constant ($d_e$) or baryon number ($g_B$). Thus, HeLIOS can simultaneously search for UDM using four different DM-SM coupling channels. Solving Eq.~\eqref{eq:DetectorNoiseBudet} for $d_{m_e}$ or $g_{B-L}$ yields the region of parameter space that HeLIOS could access with a quantum-limited transducer for scalar or vector UDM, respectively. These are shown in Fig.~\ref{fig:ParameterSpace} for the first ten normal modes of the prototype detector discussed below. Also shown are the extended regions when tuning the mode frequencies through pressurization, as well as currently existing bounds.

\section{Experimental design}

\begin{figure}[t]
    \centering
    \includegraphics[width=\linewidth]{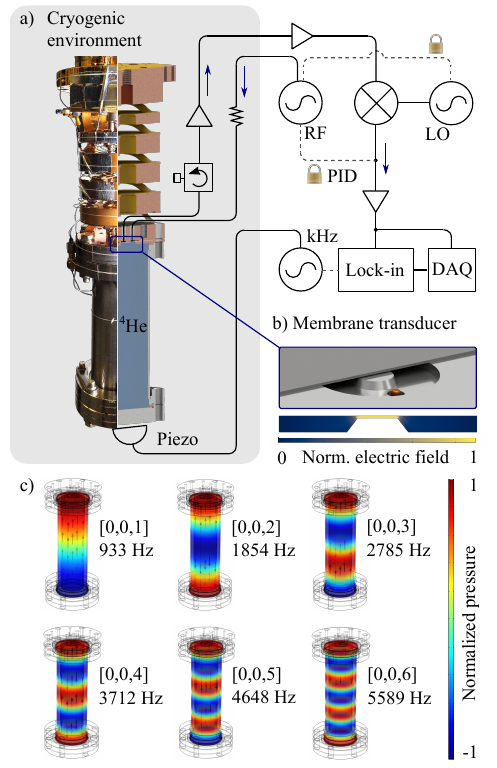}
    \caption{a) Experimental setup, including a picture and cut rendering of the detector assembly suspended from the mixing chamber of a dilution refrigerator as well as a sketch of the readout electronics. b) Rendering of the re-entrant microwave cavity transducer with a membrane as interface to the helium (cut in half for illustration purposes). c) Finite-element simulations of the lowest-order mechanical modes along the cylinder axis $[r,\varphi,z] = [0,0,n]$, with the normalized acoustic pressure fields of the helium and computed mode frequencies.}
    \label{fig:ExperimentalSetup_Simulations}
\end{figure}

A sketch of the experiment is shown in Fig.~\ref{fig:ExperimentalSetup_Simulations}. The design is similar to our previous prototype superfluid gravitational wave detector \cite{Vadakkumbatt2021}. A commercial 2.75" ConFlat nipple made of stainless steel provides a cylindrical volume for \unit[145]{ml} of superfluid helium (with \unit[12.8]{cm} length and \unit[3.8]{cm} diameter). The bottom of the cell is capped by a ConFlat flange, while the top is sealed with a niobium membrane of \unit[300]{\textmu m} thickness and \unit[1.4]{cm} diameter, clamped from below through an indium-plated copper ring to facilitate a superfluid leak-tight cell. The kilohertz acoustic modes of the helium are non-resonantly coupled to the fundamental drum mode of the membrane, whose frequency is \unit[16.2]{kHz} in the absence of helium. Mechanical dissipation of the resulting helium-membrane normal modes at millikelvin temperatures are dominated by losses in the membrane \cite{deLorenzo2017}, which were reduced through annealing and electropolishing the niobium plate \cite{Paik1976}.

The top side of the membrane forms half of a superconducting cylindrical re-entrant microwave cavity made of indium-plated copper (see Fig.~\ref{fig:ExperimentalSetup_Simulations}b), similar to the one used in Refs.~\cite{Clark2018} and \cite{Potts2020} (see the Appendix for more information). A central stub confines the electric field within a small gap to the membrane (\unit[100]{\textmu m}). The helium pressure strongly modulates the frequency of the microwave resonator through the capacitance, enabling sensitive optomechanical transduction of the helium motion. Driving the microwave cavity on resonance encodes the mechanical motion into the phase of the transmitted signal, which is amplified through a cryogenic HEMT and down-converted using a standard homodyne circuit illustrated in Fig.~\ref{fig:ExperimentalSetup_Simulations}a. A piezoelectric transducer is affixed to the bottom blank of the cell to facilitate coherent excitation of the mechanical modes.

The helium cell is suspended from the mixing chamber plate of a wet dilution refrigerator through a series of four alternating copper masses and springs to isolate the detector from mechanical vibrations \cite{deWit2019,Bignotto2005} (see the Appendix for more information). The helium fill line is thermalized through sintered heat exchangers on each stage of the dilution refrigerator, enabling a base temperature of \unit[20]{mK} as measured through a primary nuclear orientation thermometer.

\section{Characterization}

Figure \ref{fig:ExperimentalSetup_Simulations}c shows finite-element simulations of the six lowest-order acoustic pressure modes $[r,\varphi,z] = [0,0,n]$, mechanically coupled to the structural modes of the membrane and surrounding detector body. Their computed frequencies, effective masses, as well as geometrical mode overlap factors for coupling to scalar and vector UDM are shown in Table \ref{tab:ModeParameters}.

\begin{table}[t]
    \centering
    \caption{Relevant parameters of the six lowest-order longitudinal acoustic helium modes $[r,\varphi,z] = [0,0,n]$ at saturated vapor pressure, coupled to the structural deformation of the membrane and detector body. Left: results from finite-element simulations for the mechanical frequencies $f$, effective masses $\mu$ (normalized by the helium mass $M = \unit[21.0]{g}$), as well as geometrical mode overlap factors $q$ and $\beta_{12} f_{12} = \beta_{n,1} f_1 + \beta_{n,2} f_2$ for coupling to scalar and vector UDM, respectively. Right: frequencies $f$ and mechanical quality factors $Q$ measured at a temperature of \unit[20]{mK}.}
    \vspace{0.5cm}
    \begin{tabularx}{\columnwidth}{l|ccccXcc}
    & \multicolumn{4}{c}{Simulated} && \multicolumn{2}{c}{Measured}\\
    Mode & $f$ [Hz] & $\mu/M$ & $q$ [g$\,$cm] & $\beta_{12} f_{12}$ && $f$ [Hz] & $Q$ [$10^{6}$]\\
    \cmidrule{1-5}
    \cmidrule{7-8}
    $[0,0,1]$ & 933 & 0.50 & 2.97 & $4.20 \times 10^{-2}$ && 998 & 0.26\\
    $[0,0,2]$ & 1854 & 0.49 & 49.5 & $1.63 \times 10^{-6}$ && 1864 & 2.2\\
    $[0,0,3]$ & 2785 & 0.47 & 1.26 & $1.44 \times 10^{-2}$ && 2800 & 3.7\\
    $[0,0,4]$ & 3712 & 0.43 & 42.6 & $2.70 \times 10^{-6}$ && 3729 & 1.9\\
    $[0,0,5]$ & 4648 & 0.42 & 2.55 & $0.90 \times 10^{-2}$ && 4668 & 2.2\\
    $[0,0,6]$ & 5589 & 0.40 & 103 & $2.38 \times 10^{-6}$ && 5605 & 2.6\\
    \end{tabularx}
    \label{tab:ModeParameters}
\end{table}

Only the even-ordered breathing modes $n \in \{2,4,6\}$ feature a significant mode overlap factor $q_n$ for coupling to scalar UDM, with $q_6$ being the largest as a result of the spectral vicinity of $[0,0,6]$ to the structural breathing mode of the cell, which finite-element modeling shows has a frequency of \unit[5.3]{kHz}. On the other hand, only the odd-ordered modes $n \in \{1,3,5\}$ will couple to vector UDM with a substantial $\beta_{n,12} f_{12} = \beta_{n,1} f_1 + \beta_{n,2} f_2$, as a result of the differential acceleration between helium ($f_1 = -3.29 \times 10^{-2}$) and stainless steel detector body ($f_2 = 0.07 \times 10^{-2}$). Thus, HeLIOS could simultaneously search for both scalar and vector UDM, with an equal amount of susceptible modes.

\begin{figure}[t]
    \centering
    \includegraphics[width=\linewidth]{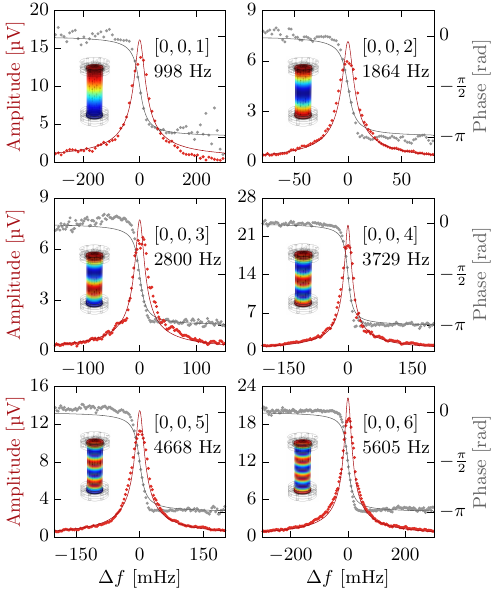}
    \caption{Lock-in amplitude and phase of the six lowest-order mechanical modes along the cylinder axis $[r,\varphi,z] = [0,0,n]$, measured at a temperature of \unit[20]{mK} and helium pressure of \unit[170]{mbar} for $[0,0,1]$ to $[0,0,3]$ as well as \unit[70]{mbar} for $[0,0,4]$ to $[0,0,6]$. The quoted frequencies are extrapolations to the helium saturated vapor pressure of $\approx~\unit[0]{mbar}$, using the linear regressions obtained in Fig.~\ref{fig:PressureTunability}.}
    \label{fig:HeliumModeSpectra}
\end{figure}

After filling the cell completely with \unit[145]{ml} superfluid helium and reaching a final base temperature of \unit[20]{mK}, the six lowest-order mechanical modes were characterized by sweeping the piezo drive frequency and coherently measuring the transmitted microwave phase with a lock-in amplifier. Fig.~\ref{fig:HeliumModeSpectra} shows the resulting amplitude and phase spectra for the modes $[0,0,1]$ to $[0,0,6]$. Only the measured mode frequency of the fundamental mode $[0,0,1]$ deviates appreciably (by \unit[7.0]{\%}) from the simulated one. This is likely a result of hybridization with another low-$Q$ mechanical mode in its spectral vicinity that could originate from the detector or suspension structure. Mechanical quality factors were obtained through ring-down measurements at a pressure of \unit[220]{mbar} and are also shown in Table \ref{tab:ModeParameters}. Low dissipation is found, with $Q$ values between 1.9 and 3.7 million, except for $[0,0,1]$ with $Q = 2.6 \times 10^5$.

\begin{figure}[t]
    \centering
    \includegraphics[width=\linewidth]{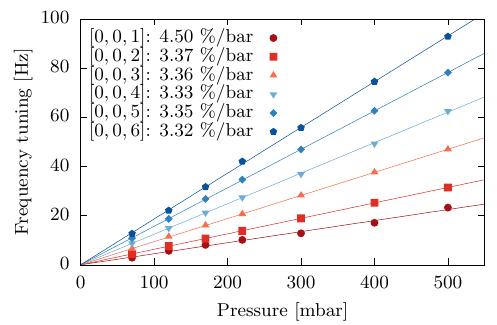}
    \caption{Frequency tuning of the six lowest-order mechanical modes along the cylinder axis $[r,\varphi,z] = [0,0,n]$ through pressurization of the helium at a temperature of \unit[20]{mK}, relative to the zero-pressure frequency $f_0$. Solid lines show linear regressions, with relative slopes (normalized to $f_0$ of each mode) quoted in the legend.}
    \label{fig:PressureTunability}
\end{figure}

Finally, the helium pressure was swept from $P = 70$ to \unit[500]{mbar} to demonstrate the frequency tunability of the mechanical modes. Fig.~\ref{fig:PressureTunability} shows the frequency shift obtained for the six lowest-order modes. The tuning in this pressure range is approximately linear, with regressions yielding slopes of \unit[$3.35 \pm 0.03$]{\%/bar} for all modes except $[0,0,1]$. These results agree well with the relative change of the helium first sound velocity in this pressure range of $\Delta c_\text{He}/\Delta P c_\text{He} = \unit[3.55]{\%/bar}$ \cite{Abraham1970,Donnelly1998}. The cell proved to be superfluid leak tight to at least \unit[7]{bar}. A maximum frequency shift of up to \unit[54]{\%} would be achievable at the helium solidification pressure of \unit[25]{bar}. The discrepancy of the $[0,0,1]$ mode in quality factor and frequency tunability is consistent with its larger frequency, supporting the assumption of an unintended mode hybridization.

In this prototype, attempts to observe thermally driven motion at \unit[20]{mK} temperature were unsuccessful. Small improvements of the transducer could be sufficient to enable thermal noise-limited readout (as discussed in the Appendix). This would facilitate displacement calibration of the time domain signal to deduce bounds on the UDM coupling constant \cite{Hauer2013}.

\section{Discussion}

Here, we have introduced the superfluid helium UDM detector HeLIOS and demonstrated its working principle. The mechanical modes could be characterized when coherently driven, featuring high quality factors and frequency tunability when pressurized. In practice, the detector could scan the accessible frequency range for the expected Doppler-broadened UDM signal shape \cite{Derevianko2018,ScalarAURIGA}. Detection protocols and false-signal tests similar to the ones used by current axion haloscope experiments like HAYSTAC \cite{HAYSTAC2020} or ADMX \cite{ADMX2021a} could be used, as their bandwidth also relies on the continuous tunability of high-$Q$ resonant modes.

Importantly, Fig.~\ref{fig:ParameterSpace} illustrates the promise of this approach for realistically achievable experimental parameters, providing a pathway to simultaneously search for dark matter via four different SM coupling channels after only an hour of integration time. After optimizing the optomechanical transduction to improve off-resonance sensitivity, future generations also have room to decrease the thermal noise floor reached on resonance. We think that a mechanical $Q$ of $10^7$, a base temperature of $\unit[10]{mK}$, and a doubled detector diameter represent realistic next-generation improvements, readily constructed on the year time scale. The dashed lines in Fig.~\ref{fig:ParameterSpace} illustrate that HeLIOS could soon be able to effectively probe unconstrained regions of UDM parameter space. As a long-term goal, the reduction of technical noise contributions could be sufficient to achieve an off-resonance sensitivity below currently existing bounds, facilitating broadband detection at any helium pressure, with a significantly wider bandwidth than the current prototype. In addition, mitigating technical and quantum noise and probing helium physics in HeLIOS will also inform other helium-based searches of both wave-like and particle-like light dark matter.

\begin{acknowledgments}
    The authors acknowledge that the land on which this work was performed is in Treaty Six Territory, the traditional territories of many First Nations, Métis, and Inuit in Alberta. They would like to thank Samy Boutros, Kripa Vyas, and Brigitte Vachon for fruitful discussions. Moreover, they acknowledge support from the University of Alberta; the Natural Sciences and Engineering Research Council, Canada (Grant Nos.~RGPIN-2022-03078, and CREATE-495446-17); the Arthur B. McDonald Canadian Astroparticle Physics Research Institute through the support of the Canada First Research Excellence Fund; the National Science Foundation Grants PHY-1912480, PHY-2047707; and the Office of the Under Secretary of Defense for Research and Engineering under award number FA9550-22-1-0323.
\end{acknowledgments}

\section*{Appendix}
\appendix*
\setcounter{equation}{0}
\setcounter{figure}{0}
\renewcommand{\theequation}{A\arabic{equation}}
\renewcommand{\thefigure}{A\arabic{figure}}

\subsection{Scalar UDM Coupling to Mechanical Resonators}

Linear coupling of scalar ultralight dark matter (UDM) to standard model (SM) particles would cause a sinusoidal modulation of the fine-structure constant $\alpha$ or fermion masses at the UDM Compton frequency $\omega_\text{dm}$ \cite{Arvanitaki2016}. Since the Bohr radius $a_0$ scales inversely proportional to $\alpha$ and the electron mass $m_e$, this results in a spatially uniform strain 
\begin{equation}
    h(t) = - d h_0 \cos(\omega_\text{dm} t)
\end{equation}
acting on a liquid or solid, with dimensionless coupling strength $d$ and amplitude $h_0 = \sqrt{8 \pi G \varrho_\text{dm}}/c \, \omega_\text{dm} \approx \unit[1.1 \times 10^{-15}]{s^{-1}}/\omega_\text{dm}$ (with local DM density $\varrho_\text{dm}$) \cite{Manley2020}. 

The strain signal leads to an amplified displacement $\bm{u}_n(\bm{x},t) = \tilde{\bm{u}}_n(\bm{x}) \xi_n(t)$ when resonantly coupled to a breathing mode $n$ of a mechanical resonator with mode shape $\tilde{\bm{u}}_n(\bm{x})$, normalized by the maximum amplitude $\xi_n(t)$. The latter can be modeled as an effective harmonic oscillator that is driven by the UDM force $F_\text{dm}(t) = q_n \Ddot{h}(t)$, with geometric mode overlap factor $q_n = \int \rho \, \tilde{\bm{u}}_n(\bm{x}) \cdot \bm{x} \, {d}^3 x$ and mass density of the detector material $\rho$ \cite{Manley2020,Hirakawa1973}. Including the finite linewidth due to the UDM coherence time $\tau_\text{dm}$, the peak power spectral density (PSD) of the scalar UDM force acting on the mode reads
\begin{equation}
    \label{eq:ScalarUDMPSD}
    S_{FF}^\text{dm}(\omega_\text{dm}) = (q_n d h_0)^2 \omega_\text{dm}^4 \tau_\text{dm}.
\end{equation}

\subsection{Vector UDM Coupling to Mechanical Resonators}

Vector UDM could couple to SM fields through the baryon number $B$ or baryon-minus-lepton number $B-L$ of an object \cite{Carney2021}. Considering the latter, a free-falling material $j$ with average proton-to-nucleon ratio $Z_j/A_j$ of its atoms will experience the acceleration 
\begin{equation}
    \tilde{\bm{a}}_j(t) = \left(1 - \frac{Z_j}{A_j}\right) g \bm{a}_0 \cos(\omega_\text{dm} t),
\end{equation}
with dimensionless coupling strength $g$ and amplitude $|\bm{a}_0| = \sqrt{2 e^2 \varrho_\text{dm}/\epsilon_0 m_\text{n}^2} \approx \unit[3.7 \times 10^{11}]{m/s^2}$ (using the nucleon mass $m_\text{n}$) \cite{Manley2021}. The center-of-mass (COM) acceleration of two bodies with masses $m_j$ reads $\bm{a}_\text{COM}(t) = [m_1 \tilde{\bm{a}}_1(t) + m_2 \tilde{\bm{a}}_2(t)]/M$ (with total mass $M = m_1 + m_2$). Consequently, the acceleration of each object in the COM frame is given through
\begin{equation}
    \bm{a}_j(t) = \tilde{\bm{a}}_j(t) - \bm{a}_\text{COM}(t) = f_j g \bm{a}_0 \cos(\omega_\text{dm} t),
\end{equation}
where we introduced the suppression factor 
\begin{eqnarray}
    \nonumber f_j &=& \frac{m_1 Z_1}{M A_1} + \frac{m_2 Z_2}{M A_2} - \frac{Z_j}{A_j}\\
    &=& \label{eq:SuppressionFactor}
    \begin{cases}
        \frac{m_2}{M} \left(\frac{Z_2}{A_2} - \frac{Z_1}{A_1}\right) & \text{for $j = 1$}\\
        \frac{m_1}{M} \left(\frac{Z_1}{A_1} - \frac{Z_2}{A_2}\right) & \text{for $j = 2$},
    \end{cases}
\end{eqnarray}
quantifying the $B-L$ mismatch of two materials to obtain a differential acceleration. In the limit $m_2 \gg m_1$, the COM of both objects coincides with the COM of the second material and the suppression factors become $f_1 \approx [(Z_2/A_2) - (Z_1/A_1)]$ and $f_2 \approx 0$, consistent with Ref. \cite{Manley2021}. 

The fraction of UDM acceleration that couples to the normal mode $n$ of a mechanical resonator with effective mass $\mu_n = \int \rho \, |\tilde{\bm{u}}_n(\bm{x})|^2 \, \text{d}^3x$ is given through the overlap integral \cite{Manley2021}
\begin{eqnarray}
    \nonumber\frac{\bm{F}_\text{dm}(t)}{\mu_n} &=& \frac{\int_{1 \cup 2} \bm{a}(t) \cdot \tilde{\bm{u}}_n(\bm{x}) \, \text{d}^3x}{\int_{1 \cup 2} |\tilde{\bm{u}}_n(\bm{x})|^2 \, \text{d}^3x}\\
    \nonumber &=& \sum_{j=1}^2 \underbrace{\frac{\int_{j} \hat{\bm{a}} \cdot \tilde{\bm{u}}_n(\bm{x}) \, \text{d}^3x}{\int_{1 \cup 2} |\tilde{\bm{u}}_n(\bm{x})|^2 \, \text{d}^3x}}_{= \beta_{n,j}} \, a_j(t)\\
    &=& \underbrace{(\beta_{n,1} f_1 + \beta_{n,2} f_2)}_{=\beta_{n,12} f_{12}} g \bm{a}_0 \cos(\omega_\text{dm} t),
\end{eqnarray}
where we defined the geometric mode overlap factors $\beta_{n,j}$. The UDM polarization $\hat{\bm{a}} = \bm{a}_0/|\bm{a}_0|$ varies over $\tau_\text{dm}$, such that it averages to $\langle a^2 \rangle \rightarrow \langle a^2 \rangle/3$ when observed with a static detector over many coherence times \cite{Manley2021}. In this case, the peak vector UDM driving force PSD becomes
\begin{equation}
    \label{eq:VectorUDMPSD}
    S_{FF}^\text{dm}(\omega_\text{dm}) = \frac{1}{3} \, \left( \mu_n \beta_{n,12} f_{12} g |\bm{a}_0| \right)^2 \tau_\text{dm}.
\end{equation}

\subsection{Detector Noise Contributions}

The respective UDM-induced driving force \eqref{eq:ScalarUDMPSD} or \eqref{eq:VectorUDMPSD} has to compete with all contributions to the detector noise. When using the acoustic modes of a test mass, thermal noise imposes a fundamental limitation regardless of the readout mechanism. The force PSD in the vicinity of a normal mode $n$ with effective mass $\mu_n$, frequency $\Omega_n$, and mechanical quality factor $Q_n$ is approximately white and reads \cite{Saulson1990}
\begin{equation}
    \label{eq:ThermalNoisePSD}
    S_{FF}^\text{th} = \frac{4 k_\text{B} T \mu_n \Omega_n}{Q_n}.
\end{equation}
For a given resonator, it can only be reduced through low temperatures $T$ and low dissipation (i.e.,~high $Q_n$). This favors the use of superfluid helium as resonant mass, featuring ultra-low dissipation limited only through three-phonon scattering and He-$3$ impurities below $\approx~\unit[600]{mK}$ \cite{Abraham1969}, with demonstrated mechanical quality factors of more than $10^8$ \cite{deLorenzo2017}.

Cavity optomechanical transducers exploit the frequency dependence of the circulating optical or microwave mode to a mechanical quantity, like the displacement of a cavity mirror \cite{Aspelmeyer2014}. Thereby, a weak signal of a mechanical mode can be parametrically upconverted, with an enhanced gain through a large photon occupation in the cavity. The displacement sensitivity is ideally limited through quantum noise --- shot noise (with displacement PSD $S_{xx}^\text{imp}$) and backaction noise (with force PSD $S_{FF}^\text{ba}$) \cite{Clerk2010}. For an overcoupled cavity with total (external) cavity decay rate $\kappa$ ($\kappa_\text{ext}$) satisfying $\kappa \approx \kappa_\text{ext} \gg \Omega_n$, these can be expressed as \cite{Schliesser2009}
\begin{equation}
    \label{eq:TechnicalNoisePSDs}
    S_{xx}^\text{imp} = \frac{\hbar^2}{S_{FF}^\text{ba}} = \frac{\kappa}{8 n_\text{c} G^2},
\end{equation}
where $n_\text{c} = 4 P_\text{c}/\kappa \hbar \omega_\text{c}$ is the average cavity photon occupation (for circulating power $P_\text{c}$ and frequency tuned to the cavity resonance $\omega_\text{c}$), and $G = \partial \omega_\text{c}/\partial x$ describes the coupling rate with mirror displacement $x$ \cite{Aspelmeyer2014}. 

Finally, the force PSD $S_{FF}^\text{noise}(\omega)$ as a function of frequency $\omega$ for all considered noise contributions \eqref{eq:ThermalNoisePSD} and \eqref{eq:TechnicalNoisePSDs} adds up to
\begin{equation}
    \label{eq:DetectorNoiseBudet_Appendix}
    S_{FF}^\text{noise}(\omega) = S_{FF}^\text{th} + |\chi(\omega)|^{-2} \, S_{xx}^\text{imp} + S_{FF}^\text{ba},
\end{equation}
using the mechanical susceptibility $\chi(\omega) = [\mu_n (\Omega_n^2 - \omega^2 + i \omega \Omega_n/Q_n)]^{-1}$.

\subsection{Microwave Optomechanical Transducer}

The cylindrical re-entrant microwave cavity of the current HeLIOS prototype has a frequency of $\omega_\text{c}/2 \pi = \unit[11.5]{GHz}$ and total cavity decay rate of $\kappa/2 \pi = \unit[11]{MHz}$ ($Q_\text{c} = \omega_\text{c}/\kappa = 1040$). When the helium applies a pressure $P$, the frequency of the microwave resonator is strongly modulated with a coupling rate of $\partial \omega_\text{c}/\partial P = - 2 \pi \times \unit[374]{MHz/bar}$. Consequently, the single-photon single-phonon coupling rate lies between $g_0 = (\partial \omega_\text{c}/\partial P) \Delta P_\text{ZP} = - 2 \pi \times \unit[1.0]{\text{\textmu}Hz}$ and $- 2 \pi \times \unit[2.5]{\text{\textmu}Hz}$ for the first six longitudinal modes, with zero point displacement amplitudes $x_\text{ZP} = \sqrt{\hbar/2 \mu_n \Omega_n} \sim \unit[10^{-19}]{m}$ or pressure fluctuations $\Delta P_\text{ZP} = \sqrt{\hbar \Omega_n/\kappa_\text{He} V_{\text{eff},n}} \sim \unit[10^{-10}]{Pa}$, helium compressibility $\kappa_\text{He} = \unit[1.2 \times 10^{-7}]{Pa^{-1}}$, and effective volume $V_{\text{eff},n}$ of the respective mode pressure field \cite{deLorenzo2016}. In the current setup, a microwave power of up to $P_\text{c} \approx \unit[1]{\text{\textmu}W}$ can be applied.

\subsection{Suspension}

\begin{figure}[t]
    \centering
    \includegraphics[width=\linewidth]{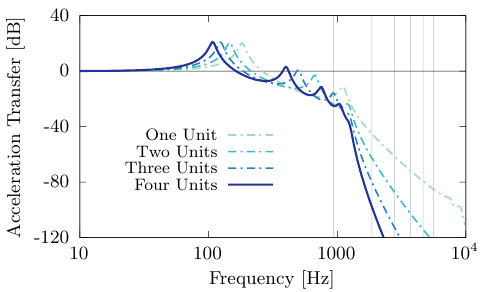}
    \caption{Finite-element simulated transfer functions of the vertical acceleration measured below the suspension with one to four mass-spring unit cells, normalized by the acceleration above the top spring. The lowest-order mechanical mode frequencies of HeLIOS are indicated as vertical lines.}
    \label{fig:Suspension}
\end{figure}

To isolate the detector from mechanical vibrations, a series of four alternating copper masses and springs suspends the helium cell and microwave transducer from the mixing chamber plate of a dilution refrigerator \cite{deWit2019,Bignotto2005}. Each mass-spring unit cell --- with a free frequency of $\approx$~\unit[60]{Hz} --- acts as a second-order mechanical low-pass filter \cite{deWit2019}, with a measured average attenuation of \unit[-16]{dB} between the \unit[-3]{dB} cutoff frequency of \unit[860]{Hz} and \unit[10]{kHz}. Fig.~\ref{fig:Suspension} shows a finite-element simulation of the acceleration transfer function through the suspension, with the number of mechanical normal modes and the steepness of the high-frequency roll-off increasing with the number of unit cells. The cantilever springs of 0.5 and \unit[0.6]{mm} thickness also facilitate thermal conduction to the detector assembly.

\subsection{Acoustic Modes in a Cylinder}

Considering only the superfluid, each acoustic pressure mode $[r,\varphi,z] = [0,0,n]$ along the cylinder axis of the detector (with length $L = \unit[12.8]{cm}$ and $z \in [-L/2, L/2]$) can be approximated as a solution to a one-dimensional wave equation, yielding
\begin{equation}
    P_n(z,t) = \cos\left[k_n \left(z + \frac{L}{2}\right)\right] \Delta P_n \sin(\Omega_n t),
\end{equation}
with acoustic pressure amplitude $\Delta P_n$ at the membrane, wavenumber $k_n = \pi n/L$, frequency $\Omega_n = 2 \pi f_n = c_\text{He} k_n$, and helium first sound velocity at saturated vapor pressure $c_\text{He} = \unit[238]{m/s}$ \cite{Donnelly1998} (leading to $f_n/n \approx \unit[930]{Hz}$). Similarly, the displacement field in the $z$-direction reads
\begin{equation}
    u_n(z,t) = \underbrace{-\sin\left[k_n \left(z + \frac{L}{2}\right)\right]}_{=\tilde{u}_n(z)} \Delta u_n \sin(\Omega_n t),
\end{equation}
with displacement amplitude $\Delta u_n$. The sinusoidal mode profile $\tilde{u}_n(z)$ would lead to a mode-independent effective mass of $\mu_n = \int \rho \, |\tilde{u}_n(z)|^2 \, \text{d}^3x = M/2$, i.e.,~equal to half of the geometrical helium mass $M = \unit[21.0]{g}$. Moreover, the geometrical mode overlap factors would be equal to 
\begin{equation}
    q_n = \int \rho \, \tilde{u}_n(z) z \, {d}^3 x =
    \begin{cases}
    0 & \text{for odd $n$}\\
    M/k_n & \text{for even $n$},
    \end{cases}
\end{equation}
for coupling to scalar UDM (with $M/k_n \approx \unit[86]{g \, cm}/n$), and
\begin{equation}
    \beta_{n,\text{He}} = \frac{\int \tilde{u}_n(z) \, \text{d}^3x}{\int |\tilde{u}_n(z)|^2 \, \text{d}^3x} = 
    \begin{cases}
    - 4/\pi n & \text{for odd $n$}\\
    0 & \text{for even $n$},
    \end{cases}
\end{equation}
for coupling to vector UDM with polarization $\hat{\bm{a}} = \hat{\bm{z}}$. Using the suppression factors as defined in Eq.~\eqref{eq:SuppressionFactor}, one obtains $\beta_{n,12} f_{12} = \sum_{j=1}^2 \beta_{n,j} f_j \approx - 4 f_\text{He}/\pi n \approx 4.28 \times 10^{-2}/n$ in the limit $m_\text{SS} \gg m_\text{He}$, i.e., assuming a heavy stainless steel cell structure.

The obtained frequencies, effective masses, and most geometrical mode overlap factors agree well with the ones computed through simulations of the entire assembly, as shown in Fig.~\ref{fig:ExperimentalSetup_Simulations}c and Tab.~\ref{tab:ModeParameters} of the main text. This underlines the fact that the modes are helium-like, i.e.,~most of the energy is stored in the acoustic modes of the superfluid. Discrepancies are the results of coupling to the membrane and detector body (particularly for $q_4$ and $q_6$, as the frequency of the cell's fundamental breathing mode is approached), as well as the three-dimensional nature of the displacement fields.

\subsection{Membrane Characterization}

After cooling the detector to \unit[30]{mK}, the free membrane motion was characterized by sweeping the piezo drive frequency and coherently measuring the transmitted microwave phase with a lock-in amplifier, revealing a fundamental drum mode with frequency $f_\text{mem} = \unit[16.2]{kHz}$. To quantify dissipation, ring-down measurements were conducted by repeatedly driving the membrane on resonance and subsequently observing the amplitude freely decaying according to $A(t) \propto \exp(-\pi f_\text{mem} t/Q)$, yielding a mechanical quality factor of $Q = 5.6 \times 10^4$.\\

\subsection{Thermal Noise-Limited Readout}

No thermally driven motion could be observed at \unit[20]{mK} temperature when applying $P_\text{c} \approx \unit[10]{nW}$ microwave power. Revisiting Eqs.~\eqref{eq:ThermalNoisePSD} and \eqref{eq:TechnicalNoisePSDs} with the cavity loss rate $\kappa/2 \pi = \unit[11]{MHz}$ and optomechanical coupling rate $|g_0|/2 \pi \lesssim \unit[2.5]{\text{\textmu}Hz}$ of the current generation, we find that the imprecision noise limit is comparable to thermal noise-driven motion on resonance (backaction is negligible). Relatively small improvements of both $\kappa$ and $g_0$ could significantly lower $S_{xx}^\text{imp} \sim \kappa^2/g_0^2$. These include optimizing the microwave cavity geometry, material, fabrication, and coupling to at least achieve $\kappa/2 \pi = \unit[2]{MHz}$ and $g_0/2 \pi = \unit[28]{\text{\textmu}Hz}$, which we demonstrated in our previous detector prototype \cite{Vadakkumbatt2021}. Other mechanical and electronic noise sources might also have to be eliminated to ensure thermal noise-limited readout close to resonance.

%

\end{document}